\documentclass[prd,twocolumn,floats,floatfix,showpacs,nofootinbib]{revtex4}
\usepackage{graphicx}
\usepackage{dcolumn}
\usepackage{bm}
\usepackage{graphics}
\usepackage{slashed}
\usepackage{amssymb}
\usepackage{natbib}
\usepackage{amsmath}
\usepackage{url}
\usepackage{color}
\newcommand{\bea}{\begin{eqnarray}}
\newcommand{\ena}{\end{eqnarray}}
\newcommand{\bean}{\begin{eqnarray*}}
\newcommand{\enan}{\end{eqnarray*}}

\newcommand{\fracc}[2]{\frac{\textstyle{#1}}{\textstyle{#2}}}

\begin{document}

\title{What is the origin of the mass of the Higgs boson?}

\author{M. Novello\footnote{M. Novello is Cesare Lattes ICRANet
Professor}}\email{novello@cbpf.br}
\author{E. Bittencourt}\email{eduhsb@cbpf.br}

\affiliation{Instituto de Cosmologia Relatividade Astrofisica ICRA -
CBPF\\ Rua Dr. Xavier Sigaud, 150, CEP 22290-180, Rio de Janeiro,
Brazil}
\pacs{98.80.Cq}
\date{\today}

\begin{abstract}
The purpose of this paper is to present a unified description of
mass generation mechanisms that have been investigated so far and
that are called the Mach and Higgs proposals. In our mechanism, gravity acts merely as a catalyst and the final
expression of the  mass depends neither on the intensity nor on the
particular properties of the gravitational field. We shall see that
these two strategies to provide mass for all bodies that operate
independently and competitively can be combined into a single unified
theoretical framework. As a consequence of this new formulation we
are able to present an answer to the question: what is the origin of
the mass of the Higgs boson?
\end{abstract}

\maketitle
\newcommand{\beq}{\begin{equation}}
\newcommand{\eeq}{\end{equation}}
\newcommand{\vare}{\varepsilon}
\newpage

\section{Introduction}

 In order to become a reliable candidate as a mechanism to generate
mass, there are three indispensable conditions that such mechanism
has to fulfil, to wit:

\begin{itemize}
\item{There must exist a universal field that interacts with
all kinds of particles;}
\item{This field must be such that its interaction with matter
breaks explicitly some symmetry that only massless particles
exhibit, e.g. the gauge freedom for vector fields or the chirality
for fermions;}
\item{There must exist a free  parameter such that different bodies
can acquire distinct values for their corresponding mass (the
spectrum of mass).}

\end{itemize}

There are only two fashionable candidates that fulfill the first
condition:

\begin{itemize}
\item{The gravitational field;}
\item{A scalar field $ \varphi.$}
\end{itemize}

The Higgs boson $ \varphi$ was postulated to couple universally with
all kinds of matter. The other candidate, gravity, is known to
couple with all forms of matter and energy and its universality is
recognized as a scientific truth. We note that after accepting
either one of these two fields as a good candidate that fulfills the
first requirement, it is not a hard job to elaborate scenarios such
that the other two conditions are satisfied too. We would like to
compare these two mechanisms and analyze the conditions under
which their strategies can be combined.

In order to simplify our analysis we will overview the generation of
mass for spinor fields. The generalization for bosons is made along
the same lines.  In both cases the origin of the mass of any body $
\mathbb{A}$ depends on its interaction with its surroundings
yielding an overall effect (described either as a scalar field---in
the case of the Higgs mechanism---or as the metric tensor of the
geometry of space-time, in the case of the gravitational origin) on
$ \mathbb{A}$ which is represented by a distribution of energy given
by the form
\begin{equation}
T_{\mu\nu} = \Lambda \, g_{\mu\nu}
\label{29julho}  \end{equation}

In the literature concerning General Relativity this form of
energy-momentum tensor is attributed to the cosmological constant
introduced by Einstein in order to be able to construct a
model for the geometry of the Universe. In the realm of quantum
field theory, such distribution is identified as the vacuum.
It is true that if one considers the Machian point of view that the inertia of a
body $ \mathbb{A} $  depends on the energy distribution of all others bodies in
the Universe, then $ \Lambda $ should be interpreted as the
cosmological constant. However this is not mandatory. The term
rest-of-the-universe concerns the environment of $ \mathbb{A}, $
that is the whole domain of influence on $ \mathbb{A} $ of the
remaining bodies in the Universe.

The idea of using a scalar field to be at the origin of the mass
appeared in the domain of high energy physics and it received the
name ``Higgs mechanism" \cite{veltman}. On the other hand, the
relationship of mass with gravity is a very old one. Such deep
connection has been emphasized in a qualitative way many times.

According to Mach, \cite{dicke} inertia is related to the global distribution of
energy of all particles in the Universe. From a historical point of
view, this idea led Einstein to the development of a new theory of
gravitation. However, the dependence of inertia on global structures
of the Universe was lost.

Otherwise, a mechanism of mass generation came from microphysics.
Indeed, the Higgs model produced an efficient scenario for
generating mass to all bodies which goes in the opposite direction
of Mach's proposal. In this mechanism, a global symmetry transforms
into a local one in the presence of vector gauge fields. Then, a
self-interaction term of an associated scalar field in its
fundamental state, represented by
$T_{\mu\nu}=L_{int}(\phi_0)g_{\mu\nu}$, appears as the vehicle which
provides mass to the gauge field.

Recently a new mechanism for generation of mass that is a
realization of Mach\rq s idea was proposed \cite{novelloCQG}. The
main idea of this new proposal is to couple nonminimally the field
under consideration to gravity through the space-time curvature. The
vacuum energy distribution of the rest-of-the-universe is
represented by the cosmological term $\Lambda$. In the realm of
General Relativity the dynamics of the metric of the space-time
together with the $\Lambda$-term is precisely responsible to
generate mass to the field.

The great novelty of this mechanism is that the gravitational field
acts merely as a catalyst, once the final expression of the mass
depends neither on the intensity nor on the particular properties of
the gravitational field. It was precisely the wrong belief that the
value of the mass obtained through any gravitational scheme should
depend on the properties of the gravitational field that was
responsible for not considering gravity as an important actor in the
mechanism of generating mass. We review briefly the alternative
Higgs mechanism in order to compare both processes.

\section{The mass of fermions}

\subsection{The Higgs vacuum $ T_{\mu\nu} = \Lambda \, g_{\mu\nu}$ }

Consider a theory of a real scalar field $ \varphi$ described by the
Lagrangian
\begin{equation}
\mathbb{L} = \frac{1}{2} \,\partial_{\mu} \varphi \, \partial^{\mu}
\varphi - V(\varphi), \label{741}
\end{equation}
where the potential has the form
$$ V = \frac{1}{2} \, \mu^{2} \, \varphi^{2} + \frac{\lambda}{4} \,
\varphi^{4}. $$ In the homogeneous case, in order to satisfy the
equation of motion, the field must be in an extremum of the
potential, which is true for two classes of solution: either
$$ \varphi = 0, $$ or
$$ \varphi_{0}^{2} = - \, \frac{\mu^{2}}{\lambda}. $$
In order to be a minimum the constant $ \mu^{2} $ must be negative.
This is a problem, since it should imply that the mass of the scalar
field is imaginary. This difficulty is in general avoided in the
following manner. One starts by redefining the field by the
introduction of a new real variable $ \chi $ :
$$ \varphi = \varphi_{0} + \chi,$$
where  $ \varphi_{0} $ is a constant. Substituting this definition
on Lagrangian (\ref{741}), it follows that
\begin{equation}
\mathbb{L} = \frac{1}{2} \,\partial_{\mu} \chi \,
\partial^{\mu} \chi + \mu^{2} \, \chi^{2} - \frac{\lambda}{4} \,
\chi^{4} - \lambda \, \varphi_{0} \, \chi^{3}  + \frac{\mu^{4}}{4
\lambda}. \label{743}
\end{equation}
This Lagrangian represents a real scalar field $ \chi $ with real
positive mass $ m^{2} = - \, \mu^{2} $ and extra terms of
self-interaction. Note that in the Lagrangian it appears a residual
constant term representing a background constant negative energy
distribution
$$ T_{\mu\nu}(residual) =- \frac{\mu^{4}}{4 \lambda} \, g_{\mu\nu}.$$

In the realm of high-energy physics it is considered that such a
term ``... has no physical consequences and can be dropped"
\cite{zumino}. We will come back to this when we analyze its
gravitational effects.

Note that now, the potential of field $ \chi $ takes the form
$$ V = m^{2} \, \chi^{2} + \frac{\lambda}{4} \, \chi^{4} + \lambda \,
\varphi_{0} \, \chi^{3}. $$ Its minimum occurs for $ \chi = 0$, which
is the point around which the expansion of the field must be made.
Let us couple this scalar field with a spinor $ \Psi$ through the
Lagrangian
\begin{equation}
\mathbb{L} = \frac{1}{2} \,\partial_{\mu} \varphi \, \partial^{\mu}
\varphi - V(\varphi) + \mathbb{L}_{D} + L_{int} \label{744}
\end{equation}
where $ \mathbb{L}_{D} $ is Dirac dynamics for massless free field.
The interaction term is $L_{int} = - \xi \, \varphi \, F(\Phi)$, where
we define $ \Phi \equiv \bar{\Psi} \, \Psi.$ Expanding in power series
with respect to $ \Phi $ and keeping the first order, as usual, we set
this term as $ - \xi \,
\varphi \, \Phi =  - \xi \, \varphi \, \bar{\Psi} \, \Psi$ . Making
the same replacement that we previously made using $ \chi$, instead of $
\varphi$, this theory becomes
\begin{equation}
\mathbb{L} = \frac{1}{2} \,\partial_{\mu} \chi \, \partial^{\mu}
\chi - V(\chi) + \mathbb{L}_{D} - \xi \, (\varphi_{0} + \chi)  \,
\bar{\Psi} \, \Psi. \label{745}
\end{equation}
The equation for the spinor field becomes
\begin{equation}
i\gamma^{\mu} \partial_{\mu} \, \Psi - \xi \, \varphi_{0} \, \Psi -
\xi \, \chi  \, \bar{\Psi} \, \Psi = 0, \label{746}
\end{equation}
which represents a spinor field of mass $ \xi \, \varphi_{0} > 0 $
interacting with a scalar field $ \chi.$

\subsection{The gravitational vacuum $ T_{\mu\nu} = \Lambda \,
g_{\mu\nu}$}

We restrict our analysis here only to the case of fermions. The
massless theory for a spinor field is given by Dirac equation
\begin{equation} i\gamma^{\mu} \partial_{\mu} \, \Psi  = 0. \label{221}
\end{equation}
This equation is invariant under $\gamma^{5} $ transformation. In
order to have mass for the fermion this symmetry must be broken. Who
is the responsible for this?

\begin{center}
\vspace{0.50cm}
\textbf{Gravity breaks the symmetry}
\vspace{0.50cm}
\end{center}
In the generation of mass through the mechanism
that we are analyzing here, gravity is responsible for breaking the
symmetry. In order to understand this we have to face the following
question: how gravity interacts with fermions? Many authors argue
that the minimal coupling principle drives this interaction. However
there are others principles involved in the choice of the form that
gravity acts on matter. Let us take for instance the example of the
scalar field. Following the paradigm of the electromagnetic field,
it is generally accepted that the dynamics of massless fields should
be invariant under conformal transformations. In the case of scalar
field this can be achieved only by taking into account a direct
coupling with the curvature of space-time. In other words, the
dynamics of the scalar field in the presence of gravity acquires a
term of the form

$$ L_{int} = \frac{1}{6} \, R \, \varphi^{2}. $$

Indeed, the presence of this extra term is sufficient to yield the
conformal invariance of the dynamics of the scalar field. We have
presented this example here only to remind the reader that the
status of the minimal coupling principle to drive the dynamics of
the interaction of any field with gravity is not a imperative law
that should be followed in any circumstance but only a suggestion
that happens to be valid for tests particle. It cannot be
transformed in a paradigm for the behavior of arbitrary fields under
 gravitational interaction.

Thus, in the framework of General Relativity we set for the
gravitational interaction of the fermion the following Lagrangian
\begin{eqnarray}
\mathbb{L} &=& \frac{i}{2} \bar{\Psi} \gamma^{\mu} \nabla_{\mu} \Psi -
\frac{i}{2} \nabla_{\mu} \bar{\Psi} \gamma^{\mu} \Psi \nonumber \\
&+& \frac{1}{\kappa} \,  (1 + \frac{\sigma}{4} \, \Phi)^{-2} \, R -
\frac{1}{\kappa}
 \, \Lambda \nonumber
\\ &-&    \, \frac{3}{\kappa} \, \frac{\sigma^{2}}{8} \,  (1 +  \frac{\sigma}{4} \, \Phi)^{-4} \, \partial_{\mu} \Phi \, \partial^{\mu} \Phi
\label{3}
\end{eqnarray}
where $\sigma$ is a constant coupling.  The first two terms of  $
\mathbb{L} $ represents the free part of the spinor field. The next
term represents the nonminimal coupling interaction of $ \Psi$ with
gravity. The vacuum---represented by $\Lambda$---is added by the
reasons presented above and it must be understood as the definition
of expressing the influence of the rest-of-the-universe on $\Psi.$ The last term of the Lagrangian is responsible to avoid higher-order derivatives of the spinor field

Using the equations of motion obtained by varying both
the metric tensor $g_{\mu\nu}$ and the spinor field $ \Psi $ and after some algebraic manipulations the equation for $\Psi $ becomes
\begin{equation} i\gamma^{\mu} \nabla_{\mu} \, \Psi  - M \Psi= 0 \label{15}
\end{equation}
where
\begin{equation}
M = \frac{\sigma \, \Lambda}{\kappa \, c^{2}}. \label{30julho13}
\end{equation}

Thus, as a result of the coupling of the spinor field with gravity
the spinor field acquires a mass $ M $ that depends crucially on the
existence of $ \Lambda.$ If $ \Lambda $ vanishes, then the mass of
the field vanishes. Let us note that there is another interpretation
of the Lagrangian (\ref{3}) that is worth pointing out here. Let us
define the nondimensional scalar field $ \eta $ by setting
$$ \eta = \frac{1}{1 + \fracc{\sigma}{4}\, \Phi}. $$
Then, in terms of this new quantity the dynamics can be rewritten
as
\begin{equation}
\mathbb{L} = \mathbb{L}_{D} - \frac{\Lambda}{\kappa} - \frac{6}{\kappa} \, (
\partial_{\mu} \eta \, \partial^{\mu} \eta - \frac{1}{6} \, R \, \eta^{2}
) \label{29junho1}
\end{equation}
which is nothing but the equation of a scalar field $ \eta $ conformally coupled to the space-time curvature.

\section{Numerical results}

Let us point out some of the observational consequences of such
mechanism. We start by recalling that the inverse Compton length of
any particle is given in terms of its mass $ M $, the Planck
constant $ \hbar $ and light velocity $ c $ yielding
$$ \mu = \frac{c}{\hbar} \, M. $$
For later use, we rewrite it in terms of
gravitational quantities using the Newton constant $ G_{N}$ or,
equivalently, the Einstein constant $ \kappa .$
The Schwarzschild solution of the gravitational field of a static
compact object has an horizon---that is a one-way membrane---characterized by its Schwarzschild radius
$$ r_{s} = \frac{1}{4\pi} \, \kappa \, M \, c^{2}.$$
Using the definition of the Planck length
$$ L_{Pl}^{2} \equiv \frac{1}{8 \pi} \, \kappa \, \hbar \, c, $$
it follows that the inverse Compton length may be written under an
equivalent form as the ratio between the corresponding Schwarzschild
radius and the Planck length squared:
\begin{equation}
\mu = \frac{1}{2} \, \frac{r_{s}}{L_{Pl}^{2}}.
\label{28jun2}
\end{equation}

The formula of the mass, obtained in Eq. (\ref{30julho13}) from the
nonminimal coupling of a spinor field $ \Psi$  with gravity, is
expressed in terms of the cosmological constant $ \Lambda ,$ the
Planck length and parameter $ \sigma $ of the nonminimal coupling
yielding the expression for the inverse Compton wavelength
\begin{equation}
\mu =  \frac{1}{8 \pi} \, \frac{\sigma \, \Lambda}{ L_{Pl}^{2}},\hspace{.3cm} \Longrightarrow \hspace{.3cm} M=\fracc{\hbar}{8\pi\, c}\,\fracc{\sigma \, \Lambda}{L_{Pl}^2}\,.
\label{28jun1}
\end{equation}
This expression relates two parameters: the mass $ M $ and the
associated nonminimal coupling constant with gravity  $ \sigma$
that has the dimensionality of volume. The knowledge of one of
these two parameters ( $ M $ or $ \sigma$ ) allows the knowledge of
its companion. By comparison of the above two expressions of  $ \mu,
$ that is, Compton definition Eq. (\ref{28jun2}) and our formula for
the mass Eq. (\ref{28jun1}) yields the expression of $ \sigma: $

\begin{equation}
\sigma = 4 \, \pi \, \frac{r_{s}}{\Lambda}. \label{28jun3}
\end{equation}

Thus, different fermionic particles that have different masses have
different values of $ \sigma.$ We note furthermore that the ratio $
M / \sigma $ which has the meaning of a density of mass is a
universal constant given only in terms of $\kappa$ and $ \Lambda.$
How to interpret such universality? There is a direct and simple way
that is the following: we rewrite this formula as a density of
energy, that is

$$ \frac{M \, c^{2}}{\sigma} = \frac{\Lambda}{\kappa}. $$

The right-hand side is nothing but the density of energy of the
vacuum. Thus we can say that  $ \sigma $ is the volume in which an
homogeneous distribution of the particle energy spreads having the
same value of the vacuum energy density provided by the cosmological
constant, that is, $ \Lambda / \kappa.$

Once our formula of mass for fermions contains gravitational
quantities which are well known to be extremely small, let us
compare it with actual numbers that we can get, for instance, from
the simplest example of the electron. The main question is: should
the coupling constant $\sigma$ become an enormously big value in
order to compensate the weakness of the gravitational field? A
direct calculation for the known elementary particles show that this
is not the case. This is a direct consequence of the fact that in
the process of giving mass gravity enters only as a catalyst. Indeed,
for the simple stable lepton, the electron $e^-$, we find that its
gravitational horizon is given by
$$ r_{s} \approx 10^{- 55} cm, $$
which implies that
$$ \sigma_{e} \approx 125 \, cm^{3}.$$

The substance that we call the electron is tremendously concentrated
within its Compton wavelength $ \lambda_{c}. $ Indeed if we compare
the density of energy $ M_{e} \, c^{2} / r^{3} $ for $
\lambda_{c}^{3} $ and $ \sigma $ it follows that all of the electron
is concentrated in the interior of its Compton volume:

$$ \frac{\varrho_{c}}{\varrho_{\sigma}} \approx 10^{31}. $$

\section{Minimal mass value}

The present method of evaluating the mass takes into account only
classical gravitational aspects. Thus, in principle it cannot be
applied at the quantum level. Indeed, quantum effects become non-negligible at least at the Compton wavelength of a given particle.
This means that there is a threshold of applicability of our
mechanism. In other words the value of the length associated to the
gravitational mechanism must be higher than the corresponding
Compton wavelength of the particle. This led naturally to the
minimum value of the mass of any fermion---called $ M_{q}$---that
can be generated by the present gravitational procedure. This value
is obtained by the condition
$$ M_{q} c^2 \geq \frac{\Lambda}{\kappa} \, \frac{\hbar^3}{M^3_{q} c^3}. $$
Inserting the current value of the constants that appear in the right-hand side
it follows that the minimum possible value for the mass is
$$ M_{q} \geq 2.36 \times 10^{-3} \, eV.$$
In the gravitational procedure of generation of mass there is no
possibility of having a fermion with a mass lower than $M_{q}.$

This procedure allows us to state that the mechanism proposed here
is to be understood as a realization of the Mach principle according to
which the inertia of a body depends on the background of the
rest-of-the-universe. This strategy can be applied in a more general
context in support of the idea that (local) properties of
microphysics may depend on the (global) properties of the Universe.
In the case $\sigma = 0$, the Lagrangian reduces to a massless
fermion satisfying Dirac\rq s dynamics plus the gravitational field
described by General Relativity.

\section{What does give mass to the scalar field that gives mass for the
vector and spinor fields?}

In the preceding section we described the Higgs model that produced
an efficient scenario for generating mass in the realm of
high-energy physics. This mechanism appeals to the intervention of a
scalar field that is the vehicle which provides mass to the spinor
field  $ \Psi.$  For the mass to be a real and constant value (a
different value for each field) the scalar field $ \varphi$ must be
in a minimum state of its potential $ V .$  This fundamental state
of the self-interacting scalar field has an low-energy distribution
given by $ T_{\mu\nu} = V(\varphi_{0}) \, g_{\mu\nu}.$ A particular
form of self-interaction of the scalar field $ \varphi$ allows the
existence of a constant value $ V(\varphi_{0}) $ that is directly
related to the mass of the original Higgs proposal.
This scalar field has its own mass, the origin of which remains unclear.

Although the concept of mass pervades almost all the analysis involving
gravitational interaction, it is an uncomfortable situation and
still to this day there has been no successful attempt to derive a
mechanism through which mass is understood as a direct consequence of a
dynamical process depending on gravity \cite{narlikar}.

According to the origins of General Relativity, the main idea concerning
inertia in the realm of gravity goes in the opposite direction
to the mechanism that we analyzed in the previous section in the
territory of high-energy physics. Indeed, while the Higgs
mechanism explores the reduction of a global symmetry into a local
one, Mach\rq s principle suggests a cosmical dependence of local
properties, making the origin of the mass of a given body dependent
on the structure of the whole universe. In this way, there ought to
exist a mechanism by means of which this quantity (mass) depends
on the state of the Universe. In precedent sections, we described a
gravitational mechanism and have shown how this vague idea can achieve a
quantitative scheme.

\subsection*{Mass for the Higgs boson}

The procedure for giving mass to bosons in our mechanism follows the
same lines. We synthesize its main steps. We start from the Lagrangian

\begin{equation}
\mathbb{L} = \frac{1}{\kappa} \, R + \frac{1}{2} \, W(\varphi) \,
\partial_{\alpha} \varphi \, \partial^{\alpha} \varphi +  B(\varphi)
\, R - \frac{1}{\kappa} \, \Lambda, \label{210}
\end{equation}

According to \cite{novelloCQG}, we choose $$ B = - \frac{\beta}{4} \,
\varphi^{2},
$$ and $$ W = \frac{2q - 6 (B')^{2}}{\alpha_{0} + 2 B}, $$

This dynamic represents a scalar field coupled nonminimally with
gravity. There is no direct interaction between $ \varphi $ and the
rest-of-the-universe (ROTU), except through the intermediary of
gravity described by a cosmological constant $ \Lambda. $ Thus $
\Lambda $ represents the whole influence of the ROTU on $\varphi.$
The dynamics of the scalar field becomes
 \begin{equation}
\Box \varphi + \mu^{2} \, \varphi = 0, \label{218}
\end{equation}
where \begin{equation}
 \mu^{2} = \beta \, \Lambda, \label{217}
\end{equation}
where, due to dimensionality arguments, we set $q = 1/ \kappa.$

Thus, as a result of the gravitational interaction, the scalar field
acquires a mass $ \mu $ that depends on the constant $\beta $ and on
the existence of $ \Lambda.$ If $ \Lambda $ vanishes then the mass
of the field vanishes. The net effect of the non-minimal coupling of
gravity with the scalar field corresponds to a specific
self-interaction of the scalar field.

\section{Combined Mach and Higgs mechanisms to generate mass}

From what we have been presenting it follows that the two mechanisms
to generate mass do not exclude each other, but instead may be
cooperative. Thus we set the complete Lagrangian as
\begin{eqnarray}
\mathbb{L} &=& \mathbb{L}_{D} + \frac{1}{2} \,\partial_{\mu} \varphi \,
\partial^{\mu} \varphi - V(\varphi) \nonumber \\
&-& \xi \, \varphi \, \bar{\Psi} \, \Psi \nonumber \\
 &-&  \frac{6}{\kappa} \, (
\partial_{\mu} \eta \, \partial^{\mu}\eta  - \frac{1}{6} \, R \, \eta^{2}) \nonumber \\
&-& \frac{\Lambda}{\kappa},
\end{eqnarray}
where we distinguish the pure Higgs term qand the pure Mach term and as
a consequence of gravitational interaction a mixed one appears.
Following the same approach applied in previous cases, we can
obtain the formula of mass that we shall analyze afterwards.

\subsection*{General mass formula} 
For an arbitrary Higgs potential, the
combined mechanism yields the mass formula

\begin{equation}
\label{mass_form_gen} 
Mc^2=\fracc{\sigma\Lambda}{k}+\fracc{\sigma
V(\phi_0)}{\hbar c}+\hbar \, c \, \xi \,\phi_0.
\end{equation}
Following the same procedure as in previous analysis and using
the standard Higgs potential, the unified Mach-Higgs mechanism
yields the corresponding expression for the mass

\begin{equation}
\label{mass_form_nu} Mc^2=\fracc{\sigma \,
\Lambda}{k}-\fracc{\sigma}{2}\fracc{m_H^2 \, c^4}{(\hbar
 \, c)^3}\nu^4+\xi \,\nu,
\end{equation}
where $\nu=\hbar \,  c \, \phi_0$.

To understand how the mass formula depends on each term we use some
values for the constants collected from Particle Data Group (\cite{pdg}), to wit:

\begin{itemize}
\item{Higgs mass range: $125\,GeV <m_Hc^2<180\,GeV$;}
\item{Cosmological constant ($\Lambda$): $1,34\times10^{-56}\,cm^{-2}$;}
\item{Einstein constant ($k$): $168,56\times10^{-39}\,\hbar c\,(GeV)^{-2}$;}
\item{Vacuum state of Higgs ($\nu$): $246\,GeV$.}
\end{itemize}

Inserting these numbers into Eq. (\ref{mass_form_nu}), we obtain

\begin{equation}
\label{mass_form_nu_numb}
\begin{array}{lcl}
Mc^2&\approx&\sigma(0,3\times10^{-2}\,\Lambda\,cm^{-1}-6,7\times10^{-6}\,cm^{-3})\\[2ex]
&&\times10^{56}MeV+246\,\xi\,GeV.
\end{array}
\end{equation}

There are several possibilities in this unified mechanism. To
exemplify these possibilities we choose the electron mass ($ M_{e}\approx 0.5\, MeV$)
to use in the calculations below. Then we list how $ M_{e}$ depends
on $ \sigma, $ $ \xi$ and $ \Lambda:$

\begin{itemize}
\item{If $\sigma=0$, the gravitational field does not play any role in the mechanism and the only surviving term is the interaction between Higgs and fermions (spontaneous symmetry breaking);}
\item{If the mechanism is dominated by Higgs-fermion interaction, then $\xi\sim10^{-6}$ and $^{3}\sqrt{\sigma}<10^{-17}\,cm$. Note that $\sigma$ is smaller than the electron Compton wavelength ($\lambda_e\approx 2.4\times 10^{-14}\,cm$);}
\item{If $\xi=0$ and $\Lambda$ assumes the cosmological value, then its contribution is very small and $^{3}\sqrt{\sigma}\sim10^{-17}\,cm$;}
\item{Whether there is no spontaneous symmetry breaking or in the absence of the Higgs boson, the unique mechanism is the gravitational one. Besides, if $\Lambda$ assumes the cosmological constant value, then $^{3}\sqrt{\sigma}\sim 5\,cm$;}
\item{In the case the value of $\Lambda$ is the order of $10^{-3}\,cm^{-2}$, then the three parts can significantly contribute for the electron mass\footnote{There is not any compelling reason to identify this constant with the actual cosmological constant or the value of the critical density $ 10^{-48} Gev^{4}$ provided by cosmology.}.}
\end{itemize}

\section{Conclusion: from Mach principle to the new gravity mechanism}

Although a widespread formulation---identified as Mach\rq s
principle---that the mass of a body may depend on the overall
properties of the rest-of-the-universe and consequently to gravity,
the association of this dependence to the smallness of gravitational
phenomena was at the origin of the general attitude of disregarding
any possibility to attribute to gravity an important role in the
generation of mass for all bodies. However, this is nothing but an
apparent difficulty and can be eliminated by two steps:

\begin{itemize}
\item{A direct coupling of matter to the curvature of space-time;}
\item{The existence of a vacuum distribution or cosmological
constant  $\Lambda.$}
\end{itemize}

This idea provides a reliable mechanism by means of which gravity is
presented as truly responsible for the generation of the mass. As a
result of such a procedure, the final expression of mass depends
neither on the intensity nor on the specific properties of the
gravitational field. This circumvents all previous criticism against
the major role of gravity in the origin of mass.

The model uses a slight modification of Mach\rq s principle. Let us
remind  that, following Einstein \cite{Einstein}, we can understand
by this principle the statement according to which the entire
inertia of a massive body is the effect of the presence of all other
masses, deriving from a kind of interaction with the latter or, in
other words, the inertial properties of a body $\mathbb{A }$ are
determined by the energy throughout all space. The simplest way to
implement this idea is to consider the state that takes into account
the whole contribution of the rest-of-the-universe onto $\mathbb{A
}$ as the most homogeneous one. Thus, it is natural to relate it to
what Einstein attributed to the cosmological constant or, in modern
language, the vacuum of all remaining bodies. This means describing
the energy-momentum distribution of all complementary bodies of
$\mathbb{A }$ in the Universe under the form
\begin{equation}
T_{\mu\nu}(U) = \frac{\Lambda}{\kappa} \, g_{\mu\nu}. \label{17abril}
\end{equation}
Note that this distribution of the energy content of the environment
of the body $\mathbb{A }$ is similar to the Higgs case, although
there is an important distinction concerning the role of this
homogeneous distribution of energy on the  generation of mass mechanism.
As we pointed out above, Higgs' proposal does not go
further to explore the consequences of this distribution of energy,
since it is not followed by the analysis relating such energy to
gravitational processes. Besides, the Higgs mechanism has a very
crucial drawback: it is obliged to assume that the mass of the Higgs
boson has a different origin than all other particles. On the other
hand, within our proposal there is a unique and the same
universal origin for the mass of all existing body. Thus, the question
``what is the origin of the mass of the Higgs boson?" has the same
answer as for all other particles, that is: the inertia of any
particle is provided by all other particles mediated by gravity that
acts as a catalyst in the fundamental vacuum---represented by the
energy distribution $ T_{\mu\nu} = \Lambda \, g_{\mu\nu} $---and as
such provides the corresponding mass. Let us make a final remark
that it is possible to understand the Mach principle in a broad sense.
Indeed, for the method of obtaining mass using the gravity
mechanism, the notion of rest-of-the-universe must be understood as
the domain of influence on a given body. As a consequence of this,
 when we deal with the vacuum represented by the distribution of energy by
$ T_{\mu\nu} = \Lambda \, g_{\mu\nu}$ it is completely irrelevant---for the gravitational mechanism of providing mass---whether parameter $\Lambda$ has a classical global origin (the Universe) identified with the cosmological constant introduced by Einstein or a restricted one (the environment) identified---as it is the case in the Higgs procedure---to the vacuum of quantum fields.

\section*{Acknowledgements}
The major part of this work was done when visiting the International
Center for Relativistic Astrophysics (ICRANet) at Pescara (Italy).
We thank professor E. Elbaz and our colleagues of ICRA-Brasil and
particularly Dr J.M. Salim for their comments in a previous version
of this paper. This work was partially supported by {\em Conselho
Nacional de Desenvolvimento Cient\'{\i}fico e Tecnol\'ogico} (CNPq),
FINEP, and {\em Funda\c{c}\~ao de Amparo \`a Pesquisa do Estado de
Rio de Janeiro} (FAPERJ) of Brazil.

\end{document}